\documentclass[a4paper,11pt]{article}
\usepackage{pos}

\graphicspath{{Figures/}}
\newcommand{\kl}{K_{\rm L}}
\newcommand{\lla}{\langle}
\newcommand{\rra}{\rangle}
\newcommand{\tsep}{t_{\rm{sep}}}
\newcommand{\chw}{\mathcal{H}_{\rm W}}
\newcommand{\gL}{\gamma^{\rm L}}

\title{$\kl\rightarrow\mu^+\mu^-$ from lattice QCD}

\author*[a,1]{En-Hung Chao}
\author[a,1]{Norman H. Christ}
\author[b,c,d]{Xu Feng}
\author[e,f,2]{Luchang Jin}

\affiliation[a]{Department of Physics, Columbia University, New York, NY 10027, USA}

\affiliation[b]{School of Physics, Peking University, Beijing 100871, China}

\affiliation[c]{Collaborative Innovation Center of Quantum Matter, Beijing 100871, China}

\affiliation[d]{Center for High Energy Physics, Peking University, Beijing 100871, China}

\affiliation[e]{Department of Physics, University of Connecticut, Storrs, CT 06269, USA}

\affiliation[f]{RIKEN-BNL Research Center, Brookhaven National Laboratory, Building 510, Upton, NY 11973}

\emailAdd{ec3693@columbia.edu}

\abstract{
We propose a lattice-QCD-suitable framework for computing the two-photon long-distance contribution to the complex $\kl\rightarrow\mu^+\mu^-$ decay amplitude, where QED is treated perturbatively in the continuum and infinite-volume.
We provide preliminary numerical results on the quark-connected diagrams on one ensemble at physical pion mass from this method, with well-controlled systematic errors.
The successful application of this method will allow the determination of the dispersive part of the aforementioned contribution from first-principles and enable a meaningful comparison between the Standard-Model prediction and experiment. 

}

\FullConference{The 40th International Symposium on Lattice Field Theory (Lattice 2023)\\
July 31st - August 4th, 2023\\
Fermi National Accelerator Laboratory\\}

\note{Supported by the U.S. D.O.E. grant \#DE-SC0011941.}
\note{Supported by D.O.E. Office of Science Early Career Award No. DE-SC0021147 and D.O.E. Award No. DE-SC0010339.}
\note{
This research used resources of the National Energy Research Scientific Computing Center (NERSC), a U.S. Department of Energy Office of Science User Facility located at Lawrence Berkeley National Laboratory, operated under Contract No.~DE-AC02-05CH11231 using NERSC award HEP-ERCAP0023253.
}

\begin{document}
\maketitle

\section{Motivation}
\begin{figure}[h!]
\centering
\includegraphics[scale=1.0]{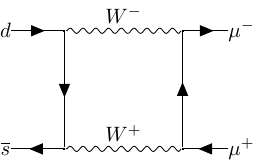}
\hspace{4pt}
\includegraphics[scale=1.0]{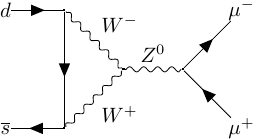}
\hspace{4pt}
\includegraphics[scale=1.0]{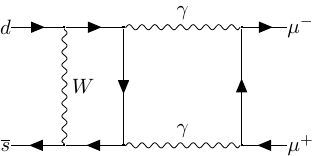}
\caption{Left and middle: short-distance contributions at one-loop. Right: long-distance two-photon contribution.}\label{fig:sdld}
\end{figure}

Experimentally, the decay rate of the $\kl\rightarrow\mu^+\mu^-$ process has been known to about 2\% since two decades~\cite{E871:2000wvm}.
On the theory side, this flavor-changing process is forbidden at tree-level in the Standard Model and is suppressed by two powers of Fermi decay constant $G_{\rm F}$ at one-loop in the electroweak expansion, which explains the smallness of the measured branching ratio of $\textrm{Br}(\kl\rightarrow\mu^+\mu^-)=6.84(11)\times 10^{-9}$~\cite{ParticleDataGroup:2022pth}.
The short-distance contributions (Fig.~\ref{fig:sdld}), which involve solely exchanges of $W$ and $Z$-bosons, have been calculated to 15\% in the Standard Model, based on perturbative methods~\cite{Gorbahn:2006bm}.
However, a full Standard-Model prediction of the decay amplitude to a comparable level of precision to the experimental result is very challenging due to the large size of the long-distance (LD) contributions, where QCD needs to be treated non-perturbatively.
Although the absorptive part of the decay amplitude from two-photon exchange can be unambiguously related to the measured process $\kl\rightarrow\gamma\gamma$, which almost saturates the observed decay rate~\cite{Cirigliano:2011ny},
the dispersive part of the LD two-photon contribution dominates the theory error budget, as it cannot be reconstructed from other experimental measurement in a systematic way
\footnote{After the presentation of the current work at the Conference, updates on the parametrization of the $\kl\rightarrow\gamma^*\gamma^*$ transition form factor based on a dispersive approach became available~\cite{Hoferichter:2023wiy, Hoid:2023has}, which would sharpen the estimate of the two-photon LD contribution to the complex $\kl\rightarrow\mu^+\mu^-$ decay amplitude from phenomenology.}.

\section{Formalism}
Similar to the lattice QCD calculations of the hadronic light-by-light contribution to the anomalous magnetic moment of the muon~\cite{Blum:2019ugy}, we expand the decay amplitude in the electroweak sector to order $G_{\rm F}\,\alpha^2_{\rm{QED}}$ and express the decay amplitude as a hadronic correlation function in coordinate space weighted by a kernel.
In Minkowski spacetime, the decay amplitude reads
\begin{equation}\label{eq:mink}
\begin{split}
\mathcal{A}_{ss'}&(k^+,k^-) =\; \left(4\pi\alpha_{\rm QED}\right)^2\int\frac{d^4 p}{(2\pi)^4} \int d^4u\; d^4v \;
   \frac{e^{-i\left(\frac{P}{2}-p\right)v}}{(\frac{P}{2}-p)^2-i\varepsilon}
 \cdot \frac{e^{-i\left(\frac{P}{2}+p\right)u}}{(\frac{P}{2}+p)^2-i\varepsilon}
\\&\times
   \frac{\overline{u}_s(k^-)\gamma_\nu\{\gamma\cdot(\frac{P}{2}+p-k^+) +m_\mu\}\gamma_\mu v_{s'}(k^+)}
      {(\frac{P}{2}+p-k^+)^2 +m_\mu^2-i\varepsilon}  \cdot \left\langle 0\left|\textrm{T}\left\{ J_\mu(u) J_\nu(v) \mathcal
{H}_W(0)\right\}\right|K_L\right\rangle\,,
\end{split}
\end{equation}
where $J_\mu(u)\equiv\sum_{f}Q_f\bar{\psi}_f\gamma_\mu\psi_f(u)$ with $Q_f$ being the charge factor is the vector current, $\mathcal{H}_W(0)$ is the effective weak Hamiltonian density, $P$ is the four-momentum for the kaon in the rest frame, $k^\pm$ are the four-momenta of the outgoing $\mu^\pm$, and $s$ and $s^\prime$ indicate the spins for the final muon pair.
Neglecting the phase of the CKM matrix elements and assuming $\kl$ to be $CP$ odd, $CP$-conservation requires the final state muon pair to be an ${}^1S_0$ state, with the two final muons having opposite helicities.

To relate the time-ordered matrix element in Eq.~\eqref{eq:mink} to an Euclidean-spacetime correlation function, computable in lattice QCD, we employed the same analytical continuation as proposed in the lattice study of $\pi^0\rightarrow e^+e^-$~\cite{Christ:2022rho}.
The idea consists in deforming the $p^0$-integration contour in the opposite direction as for the coordinate-space variables to keep the Fourier weights fixed.
Nonetheless, unlike the case of Ref.~\cite{Christ:2022rho}, we expect unphysical, exponentially growing terms to arise in our calculation due to the intermediate states which are lighter than the initial kaon~\cite{Christ:2020bzb}.
By inspecting carefully the different time orders, we can recognize two sources for this behavior:
\begin{enumerate}
\item
For the time order where the weak Hamiltonian is the earliest, the rest $\kl$ can turn into a $\pi^0$, resulting inevitably in an exponential growth as the time-separation between the Hamiltonian and the vector currents increases.
This is a common feature in some applications of lattice QCD to flavor physics~\cite{Bai:2014cva}.
Although it is necessary to identify this unphysical contribution and explicitly subtract it, it is a rather easy task that one can handle very well with state-of-the-art lattice QCD techniques.

\item 
The low-energy $\pi\pi\gamma$ intermediate states created between the two vector-currents, if allowed by finite-volume kinematics, may fail to suppress the exponential growth of the analytically-continued kernel when the two vector currents are far apart.
While it is straight-forward to remove such exponentially growing terms, in contrast to one- and two-particle states~\cite{Christ:2015pwa}, the appropriate finite-volume correction needed to compensate for the absence of such low-energy three-particle states in a finite-volume Euclidean calculation is not known.
Fortunately, these three-particle low-energy contributions are phase-space suppressed and appear safe to neglect in a calculation targeting 10\% accuracy~\cite{rbc-kll:2023}.
\end{enumerate}

In the numerical study presented in the next section, we use a lattice of linear extent 4.7 fm, where the spectral gap of a moving $\pi\pi$ state is large enough such that the resulting $\pi\pi\gamma$ intermediate state is more energetic than the rest $\kl$.
Consequently, we expect the exponential growth from the second source to be absent, for which we provide numerical evidence.

\section{Numerical implementation}
As described above, the infinite-volume Minkowski space amplitude given in Eq.~\eqref{eq:mink} can be evaluated in finite-volume, from a Euclidean-space expression of the form
\begin{equation}\label{eq:master}
\mathcal{A}(\tsep, \delta_{\rm max}, x) \equiv \sum_{\delta\leq\delta_{\rm max}} \sum_{u,v\in\Lambda} \delta_{v_0-x_0,\delta}\ e^{M_K (v_0-t_K)}\ K_{\mu\nu}(u-v)\lla J_\mu(u) J_\nu(v)\chw(x)\kl(t_K)\rra\,,
\end{equation}
where $K_{\mu\nu}(u-v)$ can be either the real or the imaginary part of the coordinate-space kernel in Eq.~\eqref{eq:mink}, and $\tsep = x-t_K$ has to be taken large enough such that the zero-momentum kaon interpolator $\kl(t_K)$ overlaps primarily with the kaon ground state.
The kernel $K_{\mu\nu}$ is symmetrized in a way that we always have $u_0\geq v_0$.
This constraint makes it sensible to define the upper bound $\delta_{\rm max}$ for the summation over the temporal separation $\delta=v_0-x_0$ between the vector current located at $v$ and the weak Hamiltonian density $\chw(x)$ (cf. Fig.~\ref{fig:dist}).
Note that Eq.~\eqref{eq:master} diverges as $\delta_{\rm max}\rightarrow+\infty$ due to the unphysical pion-intermediate-state contribution as explained previously.
The expression for this unphysical divergence is known:
\begin{equation}\label{eq:pi0}
\frac{1}{2m_\pi}\sum_{\delta\geq 1}^{\delta_{\rm max}}\sum_{u\in\Lambda}e^{(M_K-m_\pi)\delta}
\lla 0| J_\mu(u)J_\nu(v)|\pi^0\rra K_{\mu\nu}(u-v)
\lla\pi^0|\chw(v)|\kl\rra\,.
\end{equation}
which can be computed precisely on the lattice.
If the spatial extent is short enough such that the $\pi\pi$-intermediate has a large-enough spectral gap, subtracting Eq.~\eqref{eq:pi0} from Eq.~\eqref{eq:master} gives the decay amplitude in Minkowski spacetime.

For this work, we only consider two of the $\Delta S =1$ operators for the weak Hamiltonian~\cite{Buchalla:1995vs}
\begin{equation}
\chw(x) = \frac{G_{\rm F}}{\sqrt{2}}V_{us}^*V_{ud}(C_1Q_1+C_2Q_2)\,,
\end{equation}
where
\begin{equation}
Q_1\equiv (\bar{s}_a\gL_\mu d_a)(\bar{u}_b\gL_\mu u_b)\,,\quad
Q_2\equiv (\bar{s}_a\gL_\mu d_b)(\bar{u}_b\gL_\mu u_a)\,,
\end{equation}
and $C_1$ and $C_2$ are the Wilson coefficients matched to the lattice from the non-perturbative renormalization procedure~\cite{Lehner:2011fz}.

We first apply our formalism to the 24ID M\"obius Domain Wall fermion ensemble from the RBC/UKQCD collaboration~\cite{RBC:2014ntl,Tu:2020vpn}. 
The parameters of this ensemble are listed in Tab.~\ref{tab:24id}.
For this ensemble, the $\pi\pi\gamma$-intermediate state with a non-vanishing photon momentum has an energy of about 660 MeV in the rest frame of the initial kaon if the interaction between the pions is neglected; therefore, no unphysical diverging term will arise from this intermediate state. 
To verify that this is the case in full QCD with interacting pions, we calculate Eq.~\eqref{eq:master} with different cut-offs $R_{\rm max}$ on temporal separation $u_0-v_0$ applied to the kernel $K_{\mu\nu}(u-v)$.

\vspace{16pt}
\begin{minipage}{0.45\textwidth}
\begin{tabular}{c c}
\hline
\textbf{Parameter} & \textbf{Value} \\
\hline
$L^3\times T\times L_s$ & $24^3\times 64\times 24$ \\
$m_\pi$ [MeV] & 142 \\
$M_K$ [MeV] & 515 \\
$a^{-1}$ [GeV] & 1.023\\
\hline
\end{tabular}
\captionof{table}{Parameters for the 24ID ensemble used in this study. $L_s$ represents the length of the fifth dimension and $a^{-1}$ is the inverse lattice-spacing.}\label{tab:24id}
\end{minipage}
\hspace{16pt}
\begin{minipage}{0.45\textwidth}
\hspace{24pt}
\includegraphics[scale=0.45]{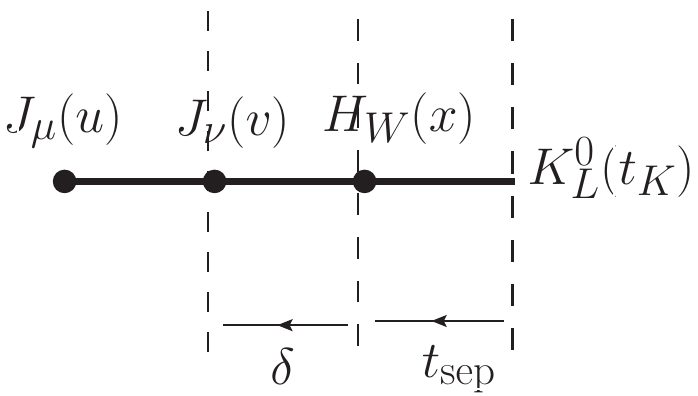}
\captionof{figure}{Definitions of $\delta$ and $\tsep$ in Eq.~\eqref{eq:master}.}\label{fig:dist}
\end{minipage}

\section{Preliminary results}
\begin{figure}[h!]
\includegraphics[scale=0.8]{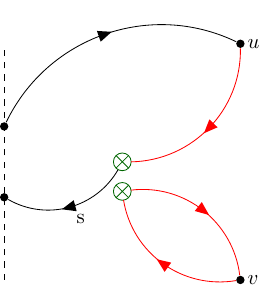}
\includegraphics[scale=0.8]{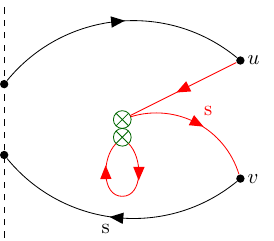}
\includegraphics[scale=0.8]{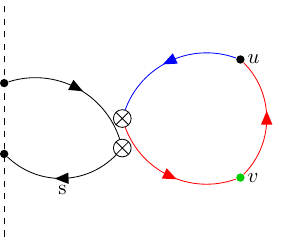}
\includegraphics[scale=0.8]{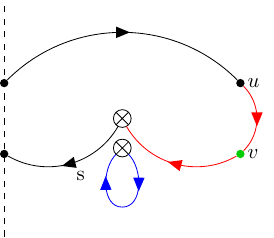}
\caption{Connected Wick-contraction topologies. From left to right: Type 1, Type 2, Type 3 and Type 4. On each diagram, the dashed line indicates the kaon wall, the crosses represent the weak Hamiltonian and the solid dots are for the vector currents. For the color code: the wall-source propagators are in black, the point-source propagators are in red, with the source positions in green, and the all-to-all ones are in blue.}
\label{fig:contr-diag}
\end{figure}

At the current stage, we have computed the four quark-connected Wick-contraction topologies listed in Fig.~\ref{fig:contr-diag}. 
Note that there are several diagrams with different flavor contents in addition to the displayed ones in Fig.~\ref{fig:contr-diag}.
A complete list of contractions can be found in Appendix E of Ref.~\cite{Zhao:2022njd}, where the first attempt at studying $\kl\rightarrow\gamma\gamma$ on the lattice has been made.
We use translational-invariance to average over measurements obtained at different source points (cf. Fig.~\ref{fig:contr-diag}) to improve the quality of the signal.
In our setup, we utilize 512 precomputed point-source propagators and 64 Coulomb-gauge-fixed wall-source propagators for both the light and strange quarks from other projects~\cite{Blum:2019ugy}.
Our code development is based on the \texttt{Grid} library~\cite{boyle:grid}.
A detailed description of the contractions will be deferred to a future publication~\cite{rbc-kll:2023}.
We will only give simplified, qualitative comments on the results on the dispersive (real) part of the amplitude and the computational strategies for the Type-1 and Type-3 diagrams in these proceedings.

\paragraph{Type 1}

\begin{figure}[h!]
\includegraphics[scale=0.35]{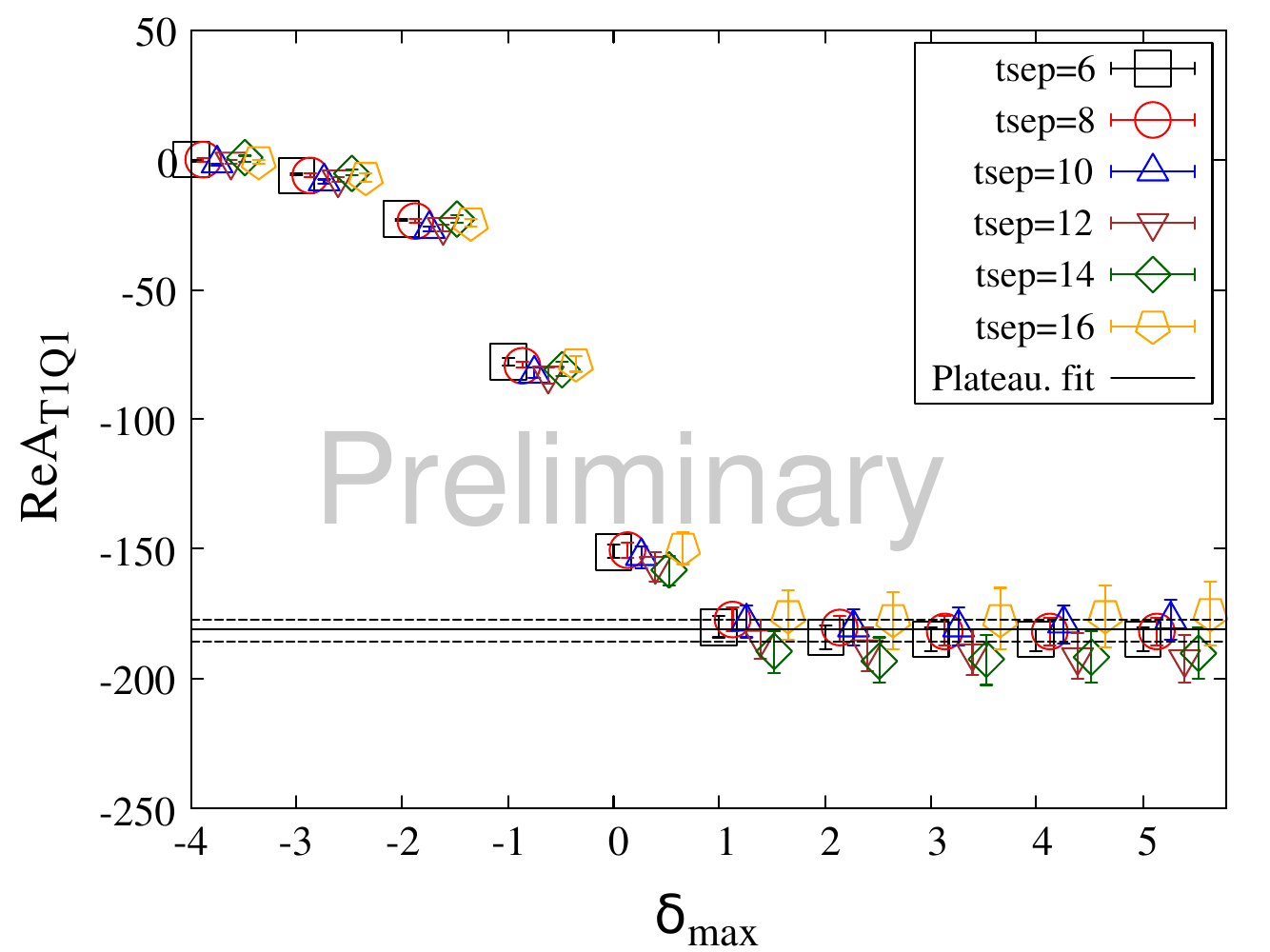}
\includegraphics[scale=0.35]{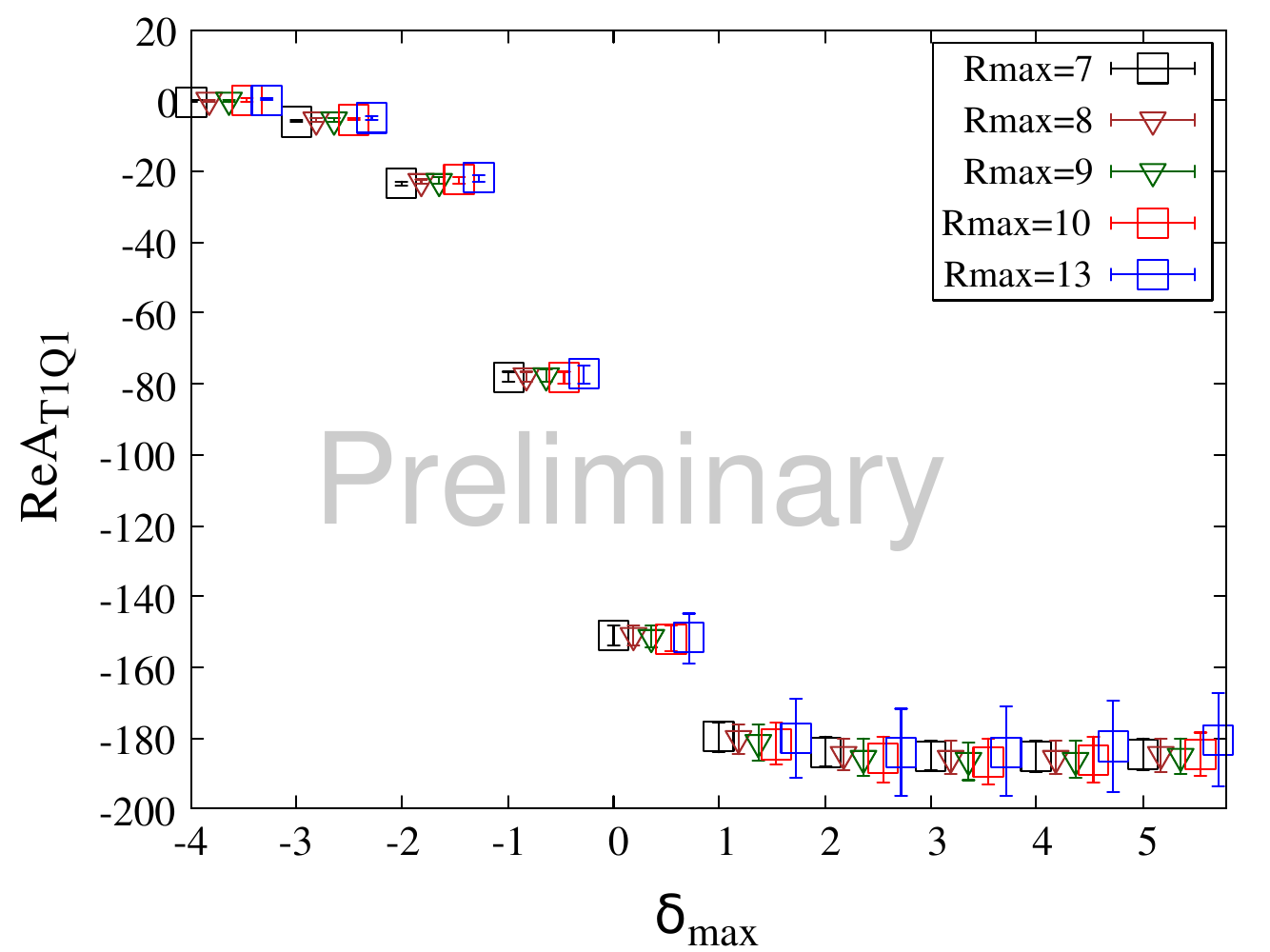}
\caption{Left: $\Re\mathcal{A}_{\rm T1Q1}$ at fixed $R_{\rm max}=7$. Right: $\Re\mathcal{A}_{\rm T1Q1}$ at fixed $\tsep=6$. The $y$-axes of both plots are in arbitrary units and the $x$-axes give the upper limit of the sum over $\delta$. The results are computed from 21 configurations.}\label{fig:t1n2}
\end{figure}

Representative preliminary results are given in Fig.~\ref{fig:t1n2}.
In the left panel, we show the contribution of one of the Type 1 diagrams to Eq.~\eqref{eq:master} at a fixed $R_{\rm max}=7$ with $\tsep=6,8,10,12,14,16$. 
The data plateau rapidly after $\delta$ goes positive, in the absence of unphysical $\pi^0$ contribution: a $\pi^0$ state can be formed only from the kaon through the weak operator; however, for Type 1 and Type 2, there are at least four quark lines between the weak operator and the subsequent vector-currents, which can only overlap with states heavier than the $\pi^0$.
The results show a nice consistency for different $\tsep$'s, which allows us to take an error-weighted average between those to have a more precise determination of the contribution. 
In the right panel of Fig.~\ref{fig:t1n2}, we plot results with different IR cut-offs $R_{\max}$ in the kernel, ranging from 7 to 13, at a fixed $\tsep=6$. 
The central values lie very nicely on top of each other, which indicates that the $\pi\pi\gamma$ state is suppressed at large distances and is too heavy to introduce an unphysical exponentially growing contribution as expected.
We could furthermore find a $R_{\max}$ value to minimize the total error, as the error increases when more distant points are included while increasing $R_{\max}$.

\begin{figure}[h!]
\includegraphics[scale=0.35]{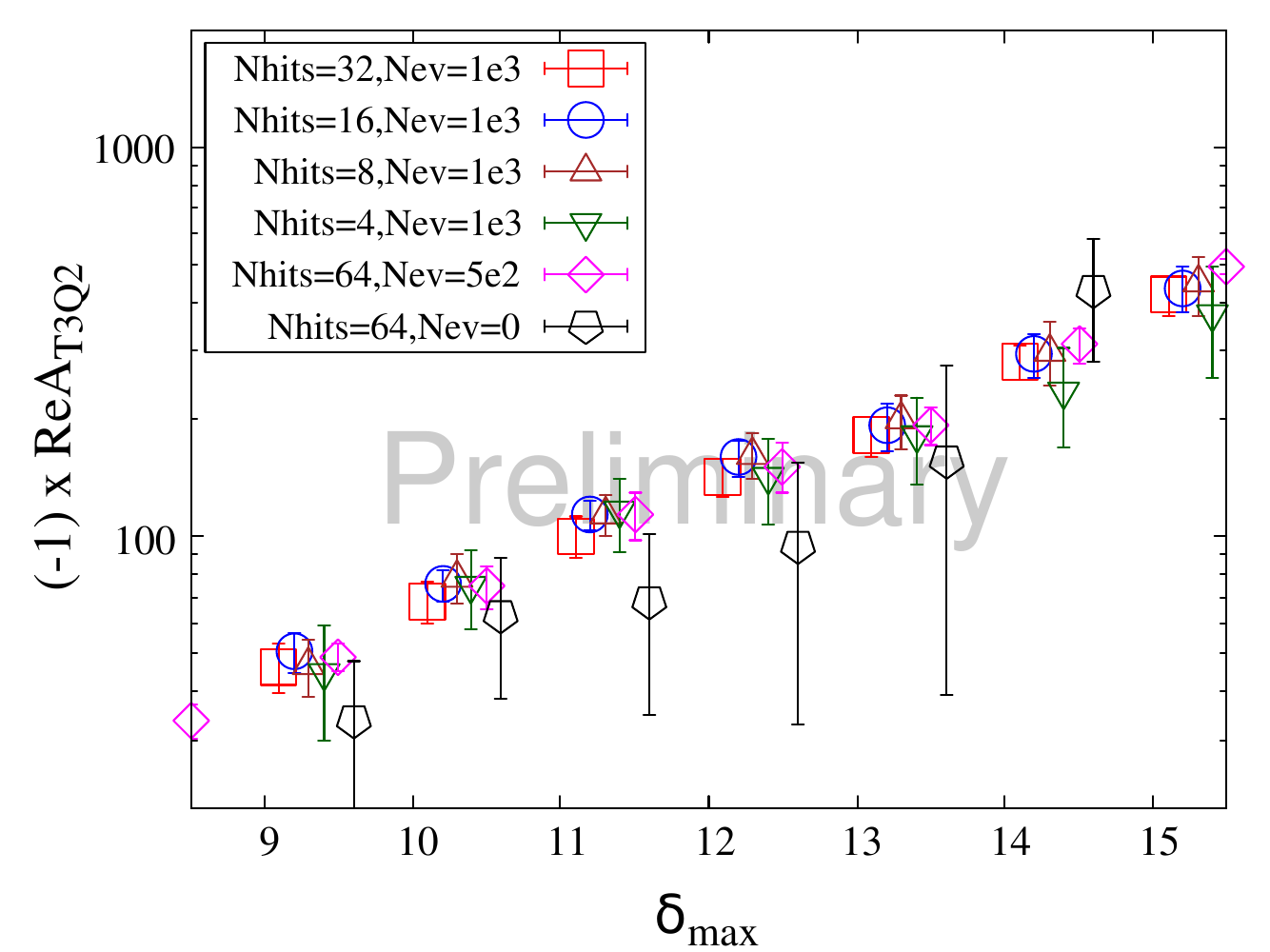}
\includegraphics[scale=0.35]{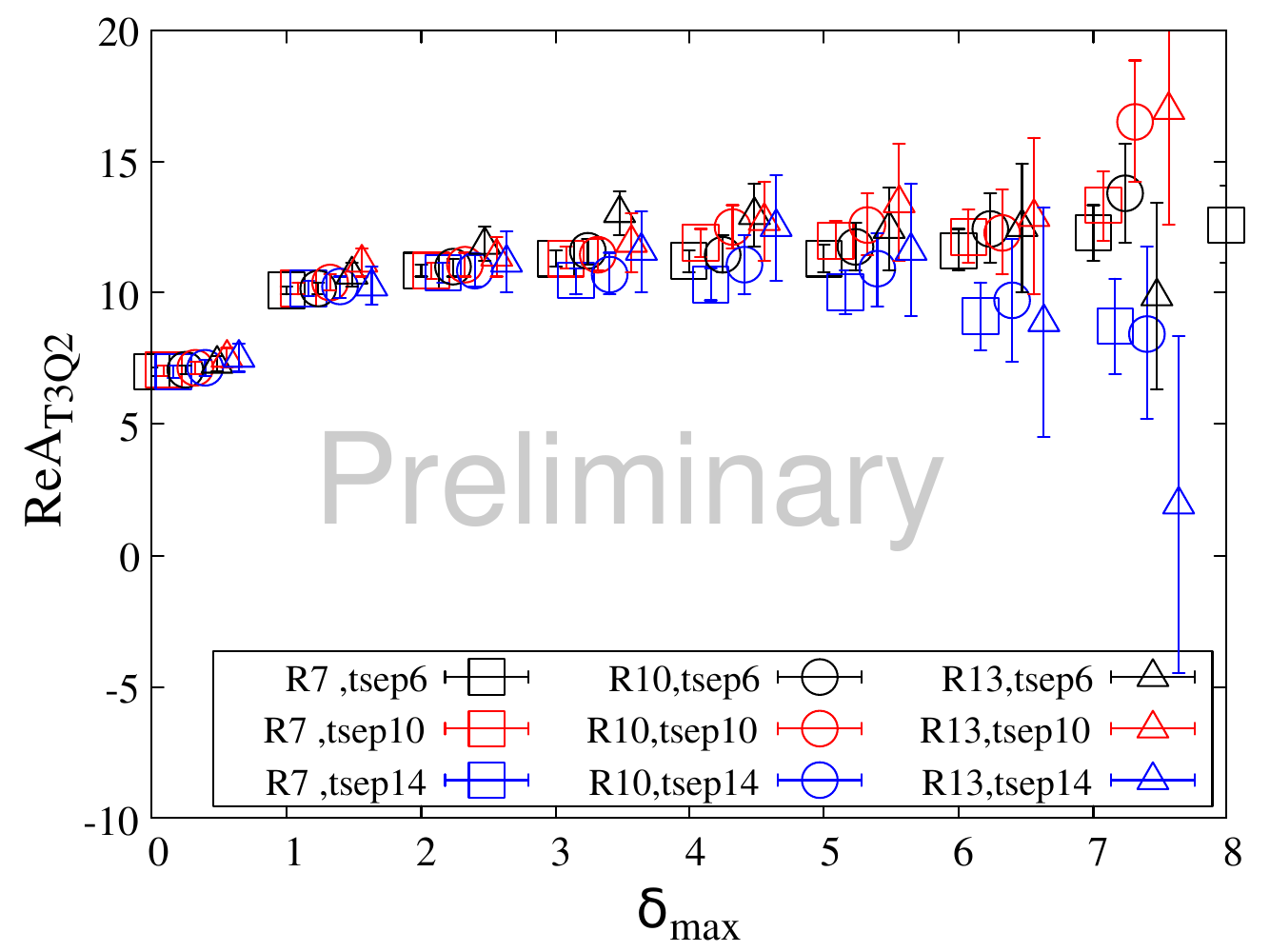}
\caption{Left: $\Re\mathcal{A}_{\rm T3Q2}$ with different numbers of low-modes (Nev) and hits. Right: $\Re\mathcal{A}_{\rm T3Q2}$ with different $\tsep$'s and $R_{\rm max}$'s after subtracting the unphysical pion intermediate state, computed with 1000 low-modes and 16 hits. The results are obtained with 110 configurations. The $y$-axes are in arbitrary units.}\label{fig:t3n4}
\end{figure}

\paragraph{Type 3}
In addition to the existing point-source and the wall-source propagator data, we have computed all-to-all propagators~\cite{Foley:2005ac} for the quark line going from the later-time current $u$ to the weak operator at $x$ for the Type-3 diagrams and for the self-contracted quark line at $x$ for the Type-4 diagrams (cf. Fig.~\ref{fig:contr-diag}).
Thanks to the available Z-M\"obius low-modes for this ensemble, we are able to decompose an all-to-all propagator into a (approximated) low-mode and a high-mode part~\cite{Mcglynn:2015uwh}.
These low-modes have originally been computed for preconditioning purposes~\cite{Yin:2011np,Clark:2017wom}.
This strategy allows us to reuse the computed propagators and convolve with several different kernels in a run to reduce overhead.
In the left panel of Fig.~\ref{fig:t3n4} we show a comparison among using different numbers of low-modes and hits to compute the contribution of the Type-3 diagram to Eq.~\eqref{eq:master} from the operator $Q_2$.
In the right panel of Fig.~\ref{fig:t3n4}, we show the result after subtracting the unphysical $\pi^0$ intermediate state Eq.~\eqref{eq:pi0}.
We extract the matrix elements appearing in Eq.~\eqref{eq:pi0} by calculating three-point functions with large-enough separations between the initial- and final-state interpolators and the rest of the operators to ensure good overlaps with the initial and final states. 
With this procedure, already at small $\delta_{\rm max}>0$, a plateau is formed, although the signal deteriorates rapidly as we go to larger $\delta_{\rm max}$.
Again, we found nice consistency among the data computed at different $R_{\rm max}$'s and $\tsep$'s.

\section{Conclusions and outlook}
In this work, a lattice-QCD-suitable framework for computing the complex $\kl\rightarrow\mu^+\mu^-$ decay rate and preliminary results on the quark-connected diagrams from its application are presented.
On a 4.7-fm lattice, we have successfully identified and subtracted the unwanted contributions, the price to pay when evaluating the Minkowski-space decay amplitude in Euclidean spacetime as required by our framework.
Ongoing efforts include the calculation of the quark-disconnected diagrams, which are expected to be noisy.
Studies on a larger physical volume to estimate finite-volume errors and multiple lattice spacings to provide a continuum limit are planned.

\section{Acknowledgement}
We thank our colleagues from the RBC/UKQCD collaboration for useful discussions and substantial technical support.

\end{document}